\pdfoutput=1 
\documentclass[a4paper,11pt]{article}
\PassOptionsToPackage{table,dvipsnames}{xcolor}
\usepackage{jheppub} 

\usepackage{dcolumn}
\usepackage{bm}
\usepackage{tikz}
\usetikzlibrary{calc}
\usetikzlibrary{patterns.meta}
\usepackage{physics}
\usepackage{cancel}
\usepackage{todonotes}
\newcommand{\Rho}{\varrho}
\newcommand{\laa}{{\lambda}}
\newcommand{\Jla}{J_{\laa}}
\newcommand{\vJla}{\vb{J}_{\laa}}

\newcommand{\jj}{j}
\newcommand{\rla}{\Rho_{\laa}}

\newcommand{\Ilaz}{I^{(z)}_\laa}
\newcommand{\Qla}{Q_{\laa}}
\newcommand{\Qlaz}{Q^{(z)}_{\laa}}
\newcommand{\Qlat}{Q^{(t)}_{\laa}}
\newcommand{\tdel}{\widetilde{\nabla}}
\newcommand{\tJla}{\widetilde{\vb{J}}_{\laa}}
\newcommand{\el}{\boldsymbol{\ell}}
\newcommand{\bsig}{\boldsymbol{\sigma}}
\newcommand{\pd}{\partial}

\newcommand{\app}{=}
\newcommand{\vcc}{}
\makeatletter  
\gdef\@fpheader{}  
\makeatother 

\begin{document}
\preprint{IPM/P-2023/17}

\title{\centering {Temporal vs Spatial\\
	Conservation and Memory Effect in Electrodynamics}}

\author[\dagger,\star]{V.~Taghiloo,}
\author[\star,\dagger]{M.H.~Vahidinia}

\affiliation[\star]{Department of Physics, Institute for Advanced Studies in Basic Sciences (IASBS),\\
	P.O. Box 45137-66731, Zanjan, Iran}
\affiliation[\dagger]{School of Physics, Institute for Research in Fundamental
Sciences (IPM),\\ P.O.Box 19395-5531, Tehran, Iran}

\emailAdd{v.taghiloo@iasbs.ac.ir, vahidinia@iasbs.ac.ir}

\abstract{We consider the standard Maxwell's theory in $1+3$ dimensions in the presence of a timelike boundary. In this context, we show that (generalized) Ampere-Maxwell’s charge appears as a Noether charge associated with the Maxwell $U (1)$ gauge symmetry which satisfies a spatial conservation equation.  Furthermore, we also introduce the notion of spatial memory field and its corresponding memory effect. Finally, similar to the temporal case through the lens of Strominger's triangle proposal, we show how spatial memory and conservation are related.}

\date{\today}
\maketitle

\section{\label{intro} Introduction}
Contrary to the thought that the gauge symmetries in gauge theories are redundancy, they can lead to non-trivial charges in the presence of \textit{boundaries} \cite{Brown:1986nw,Bondi:1962,Sachs:1962,He:2014cra,Strominger:2017zoo,Compere:2018aar}. For example, Maxwell's theory in the presence of a codimension-1 spacelike and lightlike boundaries yields an infinite extension of the global electric charge, $Q_{\laa}=\oint \laa(x) \vb{E}\cdot \dd \vb{a}$  \cite{Kapec:2015ena, He:2014cra, Strominger:2017zoo, Hosseinzadeh:2018dkh, Seraj:2016jxi, Henneaux:2018gfi, Esmaeili:2019hom, Campiglia:2017mua, Prabhu:2018gzs, Satishchandran:2019pyc}.
	This equation reduces to standard Gauss's law for $\laa(x)=1$, and we call it generalized Gauss's charge for a generic $\laa(x)$. It can be very beneficial to experiment with this charge using different surfaces. Indeed, Lorentzian geometry allows us to have three kinds of boundaries: timelike, lightlike, and spacelike.  In this paper our focus will be on timelike boundaries. These sorts of boundaries arise when we are interested in the Maxwell theory in a box \cite{PhysRevD.99.026007,Seraj:2017rzw,Esmaeili:2019mbw,Hirai:2018ijc} or as another example one may also note to the timelike boundary of AdS spacetime (for symmetries and charges of a generic timelike boundary in gravity see \cite{Adami:2022ktn}). 

Furthermore, it has been demonstrated using timelike boundaries instead of spacelike boundaries (Cauchy surfaces) provides a  more direct route to celestial holography \cite{Pasterski:2022jzc}. To be more specific, celestial currents naturally are defined on codimension-2 integrals which are spanned with a time coordinate and a spatial coordinate. The codimension-2 nature of celestial currents raises the question of whether they can be derived as surface charges of underlying gauge symmetries.  In \cite{Pasterski:2022jzc}, it was shown that the answer is affirmative and that starting with a codimension-1 timelike surface does the job.

In this paper we consider the Maxwell theory in $1+3$ dimensions in presence of a codimension-1 \emph{timelike} boundary (1+2 dimensional hypersurface).
Interestingly, we show how studying charges over these $1+2$ hypersurfaces prompts us to the generalized Ampere-Maxwell's {charge} (see equations \eqref{Q-charge-1} and \eqref{Generalized-Ampere-Max}). Similar to the generalized Gauss {charge}, its global part reduces to the standard Ampere-Maxwell law.

The definition of charges over spacelike and timelike surfaces yields two aspects of charge conservation as we call them \emph{temporal} and \emph{spatial} respectively. These definitions show up naturally in the framework of gauge theories. As a key property of charges in gauge theories, they are expressed as an integral over a codimension-2 surface and hence are called surface charges. So, they can generically depend on two coordinates and hence we can explore their conservation along each of them. Assume one of these coordinates to be time and the other one to be one of our spatial coordinates. In this regard, we can respectively define two kinds of conservation laws: temporal and spatial. We will clarify the low dimensional property of charges in gauge theories implies a neat relation between charges associated with these two types of conservation.

\textit{Memory fields} are defined as the change of a gauge
	field between early and late times when the associated field strength vanishes (see equation \eqref{DTA} and e.g. \cite{Compere:2019odm,Favata:2010zu,Afshar:2018sbq}). On the other hand, \textit{memory effects} are designed as a physical setup to relate the permanent shifts in probe quantities (such as position, velocity, and spin) to the memory fields after this long period of time \cite{Zeldovich:1974gvh,PhysRevLett.67.1486,Braginsky:1985vlg,Pasterski:2015tva,PhysRevD.98.064032,Bieri:2013hqa,Strominger:2017zoo,Pasterski:2015zua,Mao:2021eor,Susskind:2015hpa}.

In this paper, we ask if such a phenomenon also has a spatial version. Intriguingly, we will see the answer is affirmative and we define the \textit{spatial memory fields} as the change of a gauge field between two spatially distant locations where the field strength associated with the gauge field vanishes. Similar to the temporal case, we show that there exist \textit{spatial memory effects} which relate the permanent shifts in the probe quantities between two distant places.

From the Strominger triangle proposal \cite{Strominger:2017zoo}, we know that temporal conservation and memory effects are related. One can ask whether such a relation holds for spatial ones. Interestingly, we extract an explicit connection between spatial conservation and spatial memory that reveals an exact analogy with the temporal one. However, one may note that here we do not have the notion of ``soft states" and IR physics
 similar to Strominger's ``IR triangle", more precisely we instead consider ``low momentum" physics (along one of the spatial directions). \footnote{We thank M.M. Sheikh-Jabbari for bringing this to our attention.}

The organization of this paper is as follows.
In section \ref{sec:TempvsCons} we briefly review the generalized Noether charge associated with $U(1)$ gauge symmetry in the Maxwell theory and the relation between temporal conservation and Gauss's law. Then we illustrate how the spatial conservation similarly leads to generalized Ampere-Maxwell's {charge} which the correspondence charge (the generalized electric current) generates a gauge transformation. After that in section \ref{sec:Temporal-Spatial-Memory} we propose the notion of spatial memory field and its corresponding probes and explain how the memory field is generated by the generalized electric current.
Finally, in section \ref{sec:Conservation-Memory} we show how the spatial conservation of generalized Noether charge and the spatial memory effect are connected.
\section{Temporal vs Spatial Conservation} \label{sec:TempvsCons}
We consider the standard Maxwell theory of electrodynamics in the presence of a generic  matter field $\psi$ with the following action
\begin{align}\label{action}
	S[A_\alpha, \psi]=\int d^4x\qty(-\frac{1}{4}F_{\alpha \beta}F^{\alpha \beta}+\mathcal{L}_{M}(A,\psi, \pd \psi))
\end{align}
where the field strength $F_{\alpha \beta}$ is given by $F_{\alpha \beta}=\pd_{\alpha}A_{\beta}-\pd_{\beta}A_{\alpha}$. This theory respects the gauge transformation 
\begin{equation}\label{GT}
	A_{\alpha} \to A_{\alpha}+ \pd_{\alpha} \laa
\end{equation}
when is accompanied by an appropriate transformation of matter field  $\delta_{\lambda} \psi(x)$ \footnote{For example in the case of complex scalar theory $\mathcal{L}_M=(\pd _{\alpha} \psi-i e A_{\alpha}\psi)(\pd^{\alpha} \psi^{*} +i e A^{\alpha}\psi^{*})$ the transformation is $\delta_{\lambda} \psi =i{e}\laa \psi$ and $\delta_{\lambda} \psi^{*} =-i{e}\laa \psi^{*}$.}.
By varying the action with respect to the $A_{\alpha}$ and $\psi$ we can obtain the equations of motion as
\begin{equation}\label{eom}
	\pd_{\beta}F^{\alpha \beta}=j^{\alpha}, \qquad \fdv{\mathcal{L}_M}{\psi}=0,
\end{equation}
where the current $j^{\alpha}$ is defined as $j^{\alpha}=\pdv{\mathcal{L}_M}{A_{\alpha}}$. Using the standard Noether's procedure it is easy to obtain a conserved Noether current associated with the mentioned symmetry
\begin{equation}\label{Noether-current-1}
	\Jla^{\alpha}\equiv \laa j^{\alpha}+ F^{\alpha\beta} \pd_{\beta}\laa.
\end{equation}
One can simply show this Noether current is conserved on-shell, $\pd_{\alpha}\Jla^{\alpha}\app 0$. By applying the equations of motion \eqref{eom}, we can rewrite the Noether current as a total derivative
\begin{equation}\label{currentJ}
	\Jla^{\alpha}\app \pd_{\beta}(\laa(x) F^{\alpha \beta} ).
\end{equation}
As one may expect from the Noether theorem for gauge theories, the corresponding conserved charge $ \Qla=\int_{\Sigma} d \Sigma_{\alpha}  \Jla^{\alpha}$ must be given by a codimension-2 integral
\begin{align}\label{ChargeQla}
	\Qla \app \int_{\Sigma} d \Sigma_{\alpha} \pd_{\beta}(\laa(x) F^{\alpha \beta} )=\oint_{\pd\Sigma} d \Sigma_{\alpha \beta}  F^{\alpha\beta} \laa(x).
\end{align}
Here $\Sigma$ is a codimension-1 hypersurface with $\partial\Sigma$ as its boundary (boundaries)\footnote{It is useful to note that $\lambda(x)$ is a function over the whole spacetime and not just $\Sigma$ or $\partial\Sigma$.}. This is the key property of gauge theories which was mentioned in the introduction. As we will argue in the following, this equation is equivalent to the generalized form of Gauss's and Ampere-Maxwell's {charges} for \emph{spacelike} and \emph{timelike} hypersurface $\Sigma$ respectively. 

To get intuition on the Noether current \eqref{Noether-current-1}, let us look at its temporal and spatial components. To do so, we assume the metric of the spacetime to be flat $\dd s^2=-\dd t^2+h_{ij}\dd x^{i}\dd x^{j}$
\footnote{Where spatial metric $h_{ij}$ in the Cartesian, Cylindrical and Spherical coordinates takes the following form
	\begin{equation*}
		\begin{split}
			& h_{ij}\dd x^{i}\dd x^{j}=\dd x^2+\dd y^2+\dd z^2 \\
			& h_{ij}\dd x^{i}\dd x^{j}=\dd s^2+s^2 \dd\phi^2+\dd z^2 \\
			& h_{ij}\dd x^{i}\dd x^{j}=\dd r^2+r^2(\dd\theta^2+\sin^2{\theta}\dd\phi^2)
		\end{split}
\end{equation*}}
and we also use definitions $F^{0i}=E^{i}$ and $F^{ij}=\epsilon^{ijk}B_{k}$ along with $j^{\alpha}=(\rho,\vb{\jj})$ to obtain components of $\Jla^{\alpha}=(\rla,\vJla)$ as following
\begin{align}
	\rla&=\laa\,\rho +\vb{E}\cdot \grad \laa \app \div(\laa \vb{E})\, ,\\
	\vJla&=\laa \vb{\jj}-\vb{E}\pd_{t}\laa+\grad\laa\times \vb{B} \app\curl{(\laa \vb{B})} -\pd_{t}(\laa \vb{E})\, . \label{J-spatial}
\end{align}
For the global Noether current, $\laa(x)=1$, we have $\Rho_{\laa=1}=\rho\app\div \vb{E}$ and $\vb{J}_{\laa=1}=\vb{\jj}\app-\pd_{t} \vb{E}+\curl\vb{B}$. They respectively coincide with the Gauss and Ampere-Maxwell laws. Roughly speaking, we can think about the Noether current for generic $\laa(x)$ as a differential form of the generalized Gauss and Ampere-Maxwell {charges}.

In the following section, we use these results to explore well-known \textit{temporal} conservation as well as its \textit{spatial} counterpart.

\subsection{Temporal Conservation and Gauss's Law}
As mentioned, the notion of the conserved charge $\Qla$ \eqref{ChargeQla} depends on the nature of the hypersurface $\Sigma$. It is common to assume that $\Sigma=\Sigma_t$ to be a \emph{spacelike} hypersurface at $t=$\emph{const}. Then by using the relationship between  components of  strength tensor and electric field $F^{0i}=E^{i}$, the $\Qla$ reduces to (e.g \cite{Strominger:2017zoo,Kapec:2015ena,He:2014cra,Seraj:2016jxi})
\begin{align}\label{eq:Qla}
	\Qla &=\int_{\Sigma_t}\dd[3]x \rla \app \oint_{\pd \Sigma_t} \laa(x)\; \vb{E}\cdot \dd\vb{a}.
\end{align}
Simply this integral computes the flux of electric field $\vb{E}$ through the codimension-2 boundary $\pd\Sigma_t$, and is just the standard Gauss's law for  $\laa(x)=1$. For non-constant $\laa(x)$, it may be dubbed as (the integral form of) \textit{generalized} Gauss's charge.

To get more insight into the conservation of $\Qla$, it is worthwhile to calculate $\dv{\Qla}{t}$. We start from 
\begin{align}\label{time-cons}
	0& \app \pd_{\alpha}\Jla^{\alpha}= \pd_{t} \rla +\div{\vJla}
\end{align}
and perform an integration over a codimension-1 spacelike hypersurface $\Sigma_t$ to obtain
\begin{align}
	\dv{}{t} \Qla&=-\int_{\Sigma_t}d^3x\div{\vJla}=-\oint_{\pd\Sigma_t}\vJla\cdot \dd{\vb{a}}.
\end{align}
	Finally, by using the explicit form of $\vJla$ from equation \eqref{J-spatial}, we get
	\begin{equation} 
		\dv{}{t}\Qla= -\oint_{\pd\Sigma_t}(\laa\, \vb{\jj}-\vb{E} \pd_{t}\laa +\grad{\laa}\times \vb{B})\cdot \dd{\vb{a}}.
	\end{equation}
	Clearly, the rate of change in charges is given by their flux through the spacelike codimension-2 boundary $\pd \Sigma_t$ (see figure \ref{fig:hypersurfaces}). For global part, $\laa(x)=1$, we have the standard conservation of the electric charge $\dv{Q}{t}= -\oint_{\pd\Sigma_t} \vb{\jj}\cdot \dd{\vb{a}}\,$.
\tikzdeclarepattern{
	name=arrows,
	type=uncolored,
	bottom left={(-.1pt,-.1pt)},
	top right={(12.1pt,8.1pt)},
	tile size={(24pt,16pt)},
	tile transformation={rotate=0,scale=.8},
	code={
	\tikzset{x=1pt,y=1pt}
	\draw [-stealth] (0,2) -- (6,2); 
	\draw [-stealth] (6,6) -- (12,6); 
			} }
 \begin{figure}
	\begin{center}
				\begin{minipage}{.2\textwidth}
					\vspace{-.85cm}
					\begin{tikzpicture}[scale=.8]
						\coordinate (O) at (2.5,2.5);
						\coordinate (a0) at (0,0);
						\coordinate (b0) at (1,0);
						\coordinate (c0) at (2,0);
						\coordinate (d0) at (3,0);
						\coordinate (e0) at (4,0);
						\coordinate (f0) at (5,0);
						\coordinate (a1) at (0,1);
						\coordinate (b1) at (1,1);
						\coordinate (c1) at (2,1);
						\coordinate (d1) at (3,1);
						\coordinate (e1) at (4,1);
						\coordinate (f1) at (5,1);
						\coordinate (a2) at (0,2);
						\coordinate (b2) at (1,2);
						\coordinate (c2) at (2,2);
						\coordinate (d2) at (3,2);
						\coordinate (e2) at (4,2);
						\coordinate (f2) at (5,2);
						\coordinate (a3) at (0,3);
						\coordinate (b3) at (1,3);
						\coordinate (c3) at (2,3);
						\coordinate (d3) at (3,3);
						\coordinate (e3) at (4,3);
						\coordinate (f3) at (5,3);
						\coordinate (a4) at (0,4);
						\coordinate (b4) at (1,4);
						\coordinate (c4) at (2,4);
						\coordinate (d4) at (3,4);
						\coordinate (e4) at (4,4);
						\coordinate (f4) at (5,4);
						\coordinate (a5) at (0,5);
						\coordinate (b5) at (1,5);
						\coordinate (c5) at (2,5);
						\coordinate (d5) at (3,5);
						\coordinate (e5) at (4,5);
						\coordinate (f5) at (5,5);
						
						\draw[fill=Goldenrod] (a2)--(e2) node[right =.2cm,orange!50!black]{$\Sigma_{t_1}$} --(f3)--(b3)--cycle;
						\filldraw[pattern=arrows,draw=orange!50!black] (a2)--(e2)--(f3)--(b3) --cycle;

						\draw[fill=Aquamarine,draw=Blue!50!black,transform canvas={yshift = 0.5cm},opacity=.5](a2)--(e2)--(f3)--(b3)--cycle; 
						\filldraw[pattern=arrows,draw=Blue!50!black,transform canvas={yshift = 0.5cm}](a2)--(e2)--(f3)node[right =0cm,green!50!black]{$\Sigma_{t_2}$}--(b3)--cycle;

						\draw[->,opacity=.4] (O) -- ++(0:2.5cm) node[right =]{$z$};
						\draw [->,opacity=.4](O) -- ++(90:2.5cm)node[above =]{$t$};
						\draw[->,opacity=.4] (O) -- ++(225:2.5cm)node[below left =]{$x,y$};
						\draw[->,thick] ($(a2)-(0.1,-0.1)$) --  node[left] {$t$} +(90:0.5cm);
						
					\end{tikzpicture}
								
				\end{minipage}
		 		\hspace{3cm} 
		 		\begin{minipage}{.2\textwidth}
		 			
		 			\begin{tikzpicture}[scale=.8]
		 				\coordinate (O) at (2.5,2.5);
		 				\coordinate (a0) at (0,0);
		 				\coordinate (b0) at (1,0);
		 				\coordinate (c0) at (2,0);
		 				\coordinate (d0) at (3,0);
		 				\coordinate (e0) at (4,0);
		 				\coordinate (f0) at (5,0);
		 				\coordinate (a1) at (0,1);
		 				\coordinate (b1) at (1,1);
		 				\coordinate (c1) at (2,1);
		 				\coordinate (d1) at (3,1);
		 				\coordinate (e1) at (4,1);
		 				\coordinate (f1) at (5,1);
		 				\coordinate (a2) at (0,2);
		 				\coordinate (b2) at (1,2);
		 				\coordinate (c2) at (2,2);
		 				\coordinate (d2) at (3,2);
		 				\coordinate (e2) at (4,2);
		 				\coordinate (f2) at (5,2);
		 				\coordinate (a3) at (0,3);
		 				\coordinate (b3) at (1,3);
		 				\coordinate (c3) at (2,3);
		 				\coordinate (d3) at (3,3);
		 				\coordinate (e3) at (4,3);
		 				\coordinate (f3) at (5,3);
		 				\coordinate (a4) at (0,4);
		 				\coordinate (b4) at (1,4);
		 				\coordinate (c4) at (2,4);
		 				\coordinate (d4) at (3,4);
		 				\coordinate (e4) at (4,4);
		 				\coordinate (f4) at (5,4);
		 				\coordinate (a5) at (0,5);
		 				\coordinate (b5) at (1,5);
		 				\coordinate (c5) at (2,5);
		 				\coordinate (d5) at (3,5);
		 				\coordinate (e5) at (4,5);
		 				\coordinate (f5) at (5,5);
		\draw[fill=Goldenrod]($(a2)+(0.2cm,0)$)--(b2)--(c3)--($(b3)+(0.2cm,0)$)--cycle;
		 				\filldraw[pattern=arrows,draw=orange!50!black]($(a2)+(0.2cm,0)$)--(b2)--(c3)--($(b3)+(0.2cm,0)$)--cycle;

		 				\draw[fill=SkyBlue,opacity=.5]  (b0)--(c1)--(c3)--(b2)--cycle;
		 				\node[below  = .1cm of   b0,blue!50!black,,font=\large] {$\Sigma_{z_{1}}$};
		 				
		 				\draw[fill=Goldenrod] (b2)--(d2)--(e3)--(c3)--cycle;
		 				\filldraw[pattern=arrows,draw=orange!50!black](b2)--(d2)--(e3)--(c3)--cycle;
		 				
		 				\draw[fill=SkyBlue,opacity=.5]  (b2)--(c3)--(c5)--(b4)--cycle;
		 				
		 				\draw[fill=SkyBlue,opacity=.5]  (d0)--(e1)--(e3)--(d2)--cycle;
		 				\node[below  = .1cm of   d0,blue!50!black,font=\large,text opacity=1] {$\Sigma_{z_{2}}$};
		 				
		 				\draw[fill=Goldenrod] (d2)--(e2)--(f3)--(e3)--cycle;
		 				\filldraw[pattern=arrows,draw=orange!50!black] (d2)--(e2)--(f3)--(e3)--cycle;
		 				\node[right  = .2cm of   e2,orange!50!black] {$\Sigma_{t}$};
		 				\draw[fill=SkyBlue,opacity=.5]  (d2)--(e3)--(e5)--(d4)--cycle;

		 				\draw[->,opacity=.4] (O) -- ++(0:2.5cm) node[right =]{$z$};
		 				\draw [->,opacity=.4](O) -- ++(90:2.5cm)node[above =]{$t$};
		 				\draw[->,opacity=.4] (O) -- ++(225:2.5cm)node[below left =]{$x,y$};
		 				\draw[->,thick] ($(b4)+(45:.5cm)+(0:.1)$) --  node[above,near end] {$z$} +(0:1.8cm);
		 				
		 			\end{tikzpicture}		 			
					 		\end{minipage}
			\end{center}
 	\caption{Left: Electric charges in a segment of flat wire (yellow) at fixed times is given by $Q^{(t)}=\int \dd x \dd y \dd z \;\rho$.  Right: Electric charges that pass through $z=const$ sections of  a flat wire (yellow) during a time interval is given by $Q^{(z)}=\int \dd t \dd x \dd y  \; \vb{j}^{z}$.}
  \label{fig:hypersurfaces}
 \end{figure}
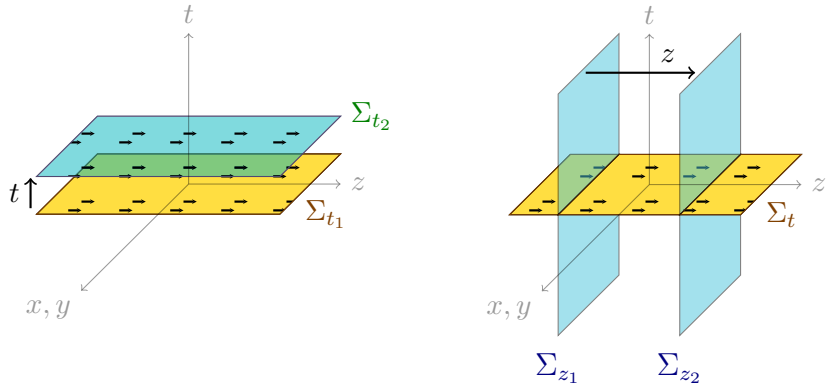
	\subsection{Spatial Conservation and Ampere-Maxwell's Law}
       As we saw in the previous subsection, the relationship between conservation and its derivation through Noether’s theorem does not rely on the codimension-one surface being spacelike (Cauchy). So it makes sense to explore this conservation for other types of surfaces as well. With this in mind, we examine Maxwell’s electromagnetism  when there is a timelike boundary present.
{In this regard,} here we assume that $\Sigma=\Sigma_z$ is a \emph{timelike} hypersurface at  $z=const$ (for reasons of simplification we will use the cylindrical coordinates system $\{t,s,\phi,z\}$).
	In this case, one may rearrange continuity equation $\pd_\alpha \Jla^{\alpha} \app 0$
	as
	\begin{align}\label{space-cons}
		0&\app \pd_{\alpha}\Jla^{\alpha}=\pd_{z}J_{\laa}^{z}+\tdel\cdot\tJla
	\end{align}
		where $\tdel\cdot$ is divergent operator on $z=const$ hypersurface (it contains derivatives along $\{t,s,\phi\}$).
	Now we can consider $J^{z}_{\laa}$ as a kind of charge density and define the corresponding charge by integration over a timelike codimension-1 hypersurface $\Sigma_z$
	as $\Qlaz=\int_{\Sigma_z}\dd[3]\widetilde{x} J^{z}_{\laa}$. One may compare this definition with $\eqref{time-cons}$. We can employ \eqref{J-spatial} to get
	\begin{align*}
				\Qlaz \app \int\dd{t}\int_{\Sigma_{zt}}\dd\bsig\cdot\Big{(}\curl{(\laa \vb{B})}- \pd_t(\laa \vb{E}) \Big{)}.
	\end{align*}
	Here $\Sigma_{zt}$ denotes a codimension-2 surface at $t,z=const$ and $d\bsig$ is its area element (in $z$ direction) \footnote{Note that in  the Cartesian coordinates $\Sigma_{zt}$ is just $x-y$ surface and $\dd\bsig=\dd{x}\dd{y}\hat{e}_z$ denotes its area element.}. 
	Now the Stokes theorem lets us to rephrase the recent equation as
	\begin{align}\label{Q-charge-1}
		\Qlaz \app \int\dd{t}\oint_{\pd \Sigma_{zt}}\hspace{-.5cm} \laa \vb{B}\cdot \dd\vb{\el} -\int\dd{t}\int_{\Sigma_{zt}}\dd\bsig\cdot\pd_{t}( \laa\vb{E}),
	\end{align} 
	where $\dd{\el}$ stands for the line element tangent to $\pd{\Sigma_{zt}}$. \footnote{The first integral in \eqref{Q-charge-1} involves a codimension-2 integral which is labeled with time and a periodic spatial coordinate. This is the direct consequence of starting with a timelike codimension-1 hypersurface. Similar codimension-2 integrals appear in celestial currents in the celestial holography context. It has been shown that these celestial currents can be derived as surface charges by starting with a timelike codimension-1 hypersurface \cite{Pasterski:2022jzc}.} 
	
	Nothing 
	$\Qlaz=\int \dd{t} \int \dd\sigma  J^{z}_{\laa}$ is just the total charge that passes hypersurface $\Sigma_{z}$ during a certain time interval, so we naturally  define a \emph{generalized electric current} as $\Ilaz\equiv \int \dd\sigma  J^{z}_{\laa} $ and hence
	\begin{equation}\label{surface-charge-general}
		\Qlaz=\int\dd{t} \Ilaz.
	\end{equation}
	Using this we get the generalized Ampere-Maxwell's {charge}
	\begin{equation} \label{Generalized-Ampere-Max}
		\boxed{  \Ilaz \app \oint_{\pd \Sigma_{zt}} \laa(x) \vb{B}\cdot \dd\vb{\el}-\int_{\Sigma_{zt}}\dd\bsig\cdot \pd_{t}(\laa(x) \vb{E}).}
	\end{equation}
	This equation reduces to the standard Ampere-Maxwell's law for $\laa(x)=1$, and $\pd_{t}(\laa(x) \vb{E})$ indicates (generalized) displacement current. Intriguingly, it implies that $\Ilaz$ as the spatial conserved charge of gauge transformation \eqref{GT}. Alternatively, a simple calculation (see appendix \ref{Symplectic Structure}) shows this charge generates gauge transformations as one may expect:
  \begin{equation} \label{PB_IA}
 \boxed{ \pb{\Ilaz}{A_{s}(s,\phi,z)}=\partial_{s} \lambda(s,\phi,z)\,.}
  \end{equation}
Equations \eqref{Generalized-Ampere-Max} and \eqref{PB_IA} are  parts of  this paper's main results; to our knowledge, they have not been previously reported.
	
	Naturally, we would like to interpret $\Ilaz$ as a conserved quantity along  spacelike direction $z$ and so introduce the notion of \emph{spatial} conservation. To illuminate this, let us start by integrating the continuity equation $\pd_{\alpha}\Jla^{\alpha}=\pd_{z}J_{\laa}^{z}+\tdel\cdot\tJla=0$ over a \emph{timelike} hypersurface $\Sigma_z$ to obtain 
	\begin{align}
		\dv{}{z} \Qlaz&=-\int_{\Sigma_z}\dd[3]\widetilde{x}\;\tdel \cdot \tJla =-\oint_{\pd\Sigma_z}\tJla \cdot \dd{\bsig}.
	\end{align}
Noteworthy, using the definition of $\Ilaz$ we can read the spatial conservation law for the generalized electric current as 
	\begin{align*}
		\Delta_{z}\Ilaz=-\int_{\Gamma}\left(\vJla+\pd_{t}(\laa \vb{E})\right)\cdot \dd \pmb{\sigma} -\Delta_z \int_{\Sigma_{zt}} \pd_{t}(\laa \vb{E}) \cdot \dd\bsig.
	\end{align*}
	Where we have defined
	$\Delta_{\vb{x}} \mathcal{O} \equiv \mathcal{O}(\vb{x}_2)-\mathcal{O}(\vb{x}_1)$ and 
	in  cylindrical coordinates $\Gamma$ denotes the lateral boundary of a cylinder.  This  shows the  change in the amount of the current between $\Sigma_{tz_2}$ and $\Sigma_{tz_1}$ surfaces  is equal to  the difference of (generalized) displacement current through them as well as the flux passing through the lateral boundary. 
	
	From Stromiger's triangle proposal \cite{Strominger:2017zoo}, conservation law (associated with asymptotic symmetries in gauge theories) and memory effects are related. So naturally one may be looking for the memory effect corresponding to the \emph{spatial} conservation law \eqref{space-cons}. By this motivation, in the following sections, we will define the \emph{spatial memory field} and its corresponding \emph{spatial memory effect} and demonstrate how this memory relates to  {spatial conservation} of \eqref{Generalized-Ampere-Max}. 
	\section{Temporal vs Spatial Memory}\label{sec:Temporal-Spatial-Memory}
	The \textit{temporal} memory field is defined as the difference of gauge potential ($A_{\alpha}\equiv(A_t,\vb{A})$) at two different times (e.g. \cite{Bieri:2013hqa,Compere:2019odm,Prabhu:2022zcr,Afshar:2018sbq})
	\begin{equation}\label{DTA}
		\Delta_{t}\vb{A}(\vb{x})=\vb{A}(t,\vb{x})\eval_{t=-T}^{t=+T}=\vb{A}(+T,\vb{x})-\vb{A}(-T,\vb{x}),
	\end{equation}
 where $\Delta_{t}\vb{A}(\vb{x})$ is a field over the space (not spacetime).	One can rewrite this equation in terms of the electric field $\vb{E}=-\pd_{t}\vb{A}+\grad{A_t}$ as follows
	\begin{equation}
		\Delta_{t}\vb{A}(\vb{x})= \int_{-T}^{T}\partial_{t} \vb{A}(t,\vb{x}) \dd t=-\int_{-T}^{T} \vb{E}(t,\vb{x}) \;\dd t.
	\end{equation}
	where the temporal gauge $A_t=0$ has been used. For simplicity's sake, let us consider the non-relativistic regime, then  Newton's second law $m\dv{\vb{v}}{t}=q\vb{E}$, implies the relationship between the temporal memory field and the change in the velocity of a charged point particle at two different times
	\begin{equation}\label{kick-memory-probe}
		\Delta_{t}{\vb{v}\eval_{\vb{x}}}=-\frac{q}{m}\Delta_{t}\vb{A}(\vb{x})\, .
	\end{equation}
 where $\Delta_{t}\vb{v}\eval_{\vb{x}}$ denotes the velocity change of a probe that initially is located at $\vb{x}$ \footnote{
 To be more precise, we note that in equation \eqref{kick-memory-probe} we have an approximation, namely,
 \begin{equation}\label{temp-approx}
		\Delta_{t}{\vb{v}\eval_{\vb{x}}}=-\frac{q}{m}\int_{-T}^{T}\partial_{t} \vb{A}(t,\vb{x}(t)) \dd t\approx-\frac{q}{m}\Delta_{t}\vb{A}(\vb{x})\, .
	\end{equation}
}.
	This relation provides a setup to measure the temporal memory effect and is dubbed as the \emph{kick memory effect} \cite{Bieri:2013hqa}. One may note that for $\vb{E}\eval_{-T}=0=\vb{E}\eval_{+T}$, the memory field  $\Delta_{t}\vb{A}(\vb{x})$ shows a pure gauge transformation which is generated by $\Qla$ \eqref{eq:Qla}. This is the first signal that memory effect and surface charges associated with large gauge transformations are related concepts \cite{Strominger:2017zoo}.
 
Nothing prevents us to define \textit{spatial} memory fields as the difference of the gauge field at two different spatial positions. For simplicity's sake, here we only consider the magnetostatics case and work in the cylindrical coordinate. In this case, we define the spatial memory field for the radial component of $\vb{A}$,
\footnote{By choosing suitable gauge conditions, one can define various kinds of spatial memory field $\Delta_{i}A_{j}$.}	
 \begin{equation}\label{DzA}
		\Delta_{z}A_{s}(s,\phi)=A_{s}(s,\phi,+Z)-A_{s}(s,\phi,-Z),
	\end{equation}
 where the memory field $\Delta_{z}A_{s}(s,\phi)$ is defined on $\Sigma_{tz}$ surface.
Similar to the temporal case, one can exploit $\vb{B}=\curl{\vb{A}}$ to express the memory field in terms of the gauge invariant quantities
\begin{equation}\label{sp-memory}
  		\boxed{  
    \Delta_{z}\vcc{A}_{s}(s,\phi)=\int_{-Z}^{+Z}\partial_{z} \vcc{A}_{s} \dd z=-\int_{-Z}^{+Z}\qty(\dd \vb{r}\times \vb{B})_{s}\, 
    }
\end{equation}
here we used the gauge $\partial_{s}\vcc{A}_{z}=0$  and assume $B_{z}=0$. Again using $m\dv{\vb{v}}{t}=q\vb{v}\times \vb{B}$, the spatial memory effect can also be expressed in terms of the velocity difference of a charged point particle
 \begin{equation}\label{spatial-memory-probe}
		\Delta_{z}v_{s}\eval_{(s,\phi)}=-\frac{q}{m}\Delta_{z}A_{s}(s,\phi),
\end{equation}
 where $\Delta_{z}v_{s}\eval_{(s,\phi)}$ shows the velocity change of a charged particle which initially is located at $(\phi,s)$ and moved from $-Z$ to $+Z$. To obtain the recent equation, we used an approximation exactly similar to the temporal case \eqref{temp-approx}.

 To clarify the notion of the spatial effect, assume we have a point particle with a non-vanishing initial velocity. In the absence of external force, this particle keeps its initial velocity. Now suppose this particle enters a region where the magnetic field is a non-zero, magnetic region, and then leaves that region. Now we can compare its initial velocity (before entering the magnetic region) and final velocity (after leaving the magnetic region). Equation \eqref{spatial-memory-probe} shows this change of velocity is encoded in the spatial memory field. It is comparable with the temporal memory effect where the interaction of a charged particle with the electromagnetic field during a finite time interval changes particle velocity \cite{Bieri:2013hqa,Pasterski:2015zua}.

	One may note that the probe of memory effect is not unique. For example, we can consider the total torque on a series of  magnetic dipoles which are located over a certain path, namely $\int (\vb{m}\times \vb{B})_x \dd z = -\vcc{m}_{z} \Delta_{z}A_{x}$, in $\partial_x A_z=0$ gauge. Here, the total torque is an observable quantity and is proportional to the memory field.

 To see how the spatial memory field and generalized conserved charge $\Ilaz$ \eqref{surface-charge-general}  are related, let us assume the magnetic field vanishes at $z=\pm Z$, it implies the gauge field $\vb{A}$ is a pure gauge at these points. Hence, the memory field $\Delta_{z}A_s(s,\phi)$ is given by a pure gauge transformation with gauge parameter $\laa'(s,\phi,+Z)-\laa(s,\phi,-Z)$. As we have discussed, in the previous section and (appendix \ref{Symplectic Structure}), this transformation is generated by $\Ilaz$ \eqref{PB_IA}.
 
As the last comment in this section, we note that the relativistic Lorentz force is given by $m\dv{\vb{p}}{t}=q(\vb{E}+\frac{1}{c}\vb{v}\times \vb{B})$ where  $p^{\alpha}=(p^{0},\vb{p})=m(U^{0},\vb{U})$ and $U^{\alpha}=\dv{x^{\alpha}}{\tau}$. By integrating over time, the first term in the Lorentz law leads to the standard temporal memory effect, and also the second term yields to the spatial memory effect $\Delta \vb{p}=-q\qty(\Delta_{t}\vb{A}+\frac{1}{c} \Delta_{\vb{x}}\vb{A})$. 	In this regard, due to the factor $1/c$, spatial memory appears as the subleading memory effect. Nevertheless, it is the leading term in the magnetostatics regime where there is no temporal memory effect.
	
	In what follow we derive an explicit relation between the spatial memory field and spatial conservation law in the magnetostatics regime.
	\section{Conservation and Memory}\label{sec:Conservation-Memory}
	In this section we explore the relation between charge conservation and memory effects \cite{Strominger:2017zoo,Pasterski:2015zua,Mao:2021eor,Susskind:2015hpa}. In this part, we only focus on the magnetostatics case. To do so, we start from the conservation of $\Ilaz$ \eqref{Generalized-Ampere-Max} in the $z$ direction
	\begin{align} \label{eq:DeltaIz}
		\Delta_{z}\Ilaz&=\Delta_z \oint \laa(x)  \vb{B}\cdot \dd{\el}.
	\end{align}
	Equivalently, we can use Stokes' theorem to write this as 
	\begin{equation}
		\Delta_{z}\Ilaz=\int \dd z\oint s \dd\phi (\vcc{B}_{\phi} \partial_{z}\laa +\laa \partial_{z}\vcc{B}_{\phi}).
	\end{equation}
	Now we restrict ourselves to the following class of gauge transformations
	\begin{equation}\label{gauge-trans-1}
		\partial_{z}\laa=\lambda_1 (\phi, s) \hspace{.2 cm} \rightarrow \hspace{.2 cm}  \laa=\lambda_{1}(\phi,s)z+\lambda_2 (\phi,s).
	\end{equation}
	Then, we get 
	\begin{equation}
		\begin{split}
			\Delta_{z}\Ilaz=&\int \dd{z}\oint s \dd\phi \lambda_1 (\partial_{z}\vcc{A}_{s}-\partial_{s}\vcc{A}_{z})
			+\int \dd{z}\oint s \dd\phi \laa \partial_{z}\vcc{B}_{\phi}
		\end{split}
	\end{equation}
	By doing the $z$-integration of the first term in the first integral and employing the equations of motion $\curl{\vb{B}}=\vb{J}$, we find
	\begin{equation*}
		\begin{split}
			\Delta_{z}\Ilaz=& \oint s \dd \phi \lambda_1 \Delta_{z} \vcc{A}_{s}-\int \dd z\oint s \dd \phi \laa \vb{\jj}_{s}\\
			-& \int \dd z\oint s \dd \phi \lambda_1 \partial_{s}\vcc{A}_{z}+\int \dd z\oint s \dd \phi \laa \partial_{\phi}\vcc{B}_{z}
		\end{split}
	\end{equation*}
 Now, we remind that the memory field \eqref{sp-memory} is defined for $B_{z}=0$ and $\partial_{s}A_z=0$ so the second line vanishes.
This restricts the allowed gauge transformations \eqref{gauge-trans-1} to
\begin{equation}
    \partial_{s}\partial_{z}\laa=0 \hspace{.2 cm} \rightarrow \hspace{.2 cm} \laa=\lambda_1(\phi)z+\lambda_2(\phi,s).
\end{equation}
After applying these conditions the final result is given by
	\begin{equation}\label{cons-memory-ems}
		\boxed{   \Delta_{z}\Ilaz= -\mathcal{I}(\laa)+\oint_{S^1} s \dd \phi\, \partial_{z} \laa \, \Delta_{z} \vcc{A}_{s}}
	\end{equation}
	where $\mathcal{I}(\laa)=\int_{\Gamma} \laa\; \vb{\jj}\cdot  \dd \bsig$. Intriguingly, this equation relates  the spatial memory field to the spatial conservation law. In particular, the spatial non-conservation of $\Ilaz$ is encoded in the flux of (hard) current, $\laa \vb{\jj}$, through the lateral boundary and memory field $\Delta_{z}\vcc{A}_{s}$. It is another main result of this paper.
	
	One can compare \eqref{cons-memory-ems} with the more familiar relation for temporal memory and temporal conservation at null infinity \cite{Prabhu:2022zcr}
	\begin{equation}
		\Delta_{u} \Qla= -\mathcal{J}(\laa)+\int_{S^2}\; \dd \Omega \; D^B\laa(\Omega) \; \Delta_{u} A_{B},
	\end{equation}
	where $\mathcal{J}(\laa)=\int_{-\infty}^{\infty}\dd u\oint_{S^2}\dd \Omega \laa(\Omega)J_{u}$ and $D_{A}$ is the covariant derivative on two sphere $S^2$.
	
	In this sense, we generalize the well-known relation between \emph{temporal} memory effect and conservation \cite{Strominger:2017zoo} to the \emph{spatial} case.
	\section{Summery and Discussion}
	\begin{figure}[t]
		\begin{center}
		\begin{tikzpicture}[scale=.8]
			\fill[
			color=Aquamarine] (0,5) circle (2cm and 0.5cm); 
			\draw[color=Blue, densely dashed,ultra thick] (-2,5) arc (180:-180:2cm and 0.5cm);
			\node[font=\large] at (0,5) {$\Sigma_t{_{+}}$};
			\draw[<-,thick,color=Blue] (2.1,5) .. controls +(.5,0) and +(-.25,-.25) .. +(.7,-.7) ; 
			\node[font=\large,color=Blue]  at (3.6,4.3){$\partial\Sigma_t{_{+}}$};
			
			\fill[ 
			color=BlueGreen] (0,0) circle (2cm and 0.5cm);
			
			\draw[color=Blue, densely dashed,ultra thick] (-2,0) arc (180:-180:2cm and 0.5cm);
			\node[font=\large] at (0,0) {$\Sigma_{t_{-}}$};
			\draw[<-,thick,color=Blue] (2.1,0) .. controls +(.5,0) and +(-.25,.15) .. +(.7,.8) ; 
			\node[color=Blue,font=\large]  at (3.6,.9){$\partial\Sigma_{t_{-}}$};
			
			\fill[
			color=Goldenrod] (0,2.5) circle (2cm and 0.5cm); 
			\draw[densely dashed,ultra thick,color=Red] (-2,2.5) arc (180:-180:2cm and 0.5cm);
			\node[] at (0,2.5) {$\Sigma_t$};
			\draw[<-,thick,color=Red] (2.1,2.5) -- +(.7,0) ; 
			\node[color=Red,font=\large]  at (3.5,2.5){$\partial\Sigma_t$};					
			
			\fill[left color=gray!50!black,right color=gray!50!black,middle color=gray!50,shading=axis,opacity=0.13,color=Blue] (2,0) -- (2,5) arc (360:180:2cm and 0.5cm) -- (-2,0) arc (180:360:2cm and 0.5cm);
			
			\draw[color=black] (-2,5) -- (-2,0);

			\draw[->,thick] (-2.8,2) .. controls +(0,+.2) and +(-.25,-.15) .. +(.7,1.5) ; 
			\node[color=black,font=\Large]  at (-2.8,1.6){$\Sigma_{r}$};
			
			\draw[->,thick] (-2.8,3.3)--(-2.8,4.3); \node[color=black,font=\Large] at (-3.1,4.2) {$t$}; 
		\end{tikzpicture}
		\caption{$\Sigma_{t}$ shows a spacelike hypersurface ($t=const$) with $\partial \Sigma_{t}$ boundary. $\Sigma_{r}$ denotes $r=const$ timelike hypersurface. Note that boundary of $\Sigma_{r}$ is disconnected $\partial \Sigma_r=\partial \Sigma_{t_+} \cup \partial \Sigma_{t_-}$.}\label{fig:stcharges}
		\end{center}
	\end{figure}
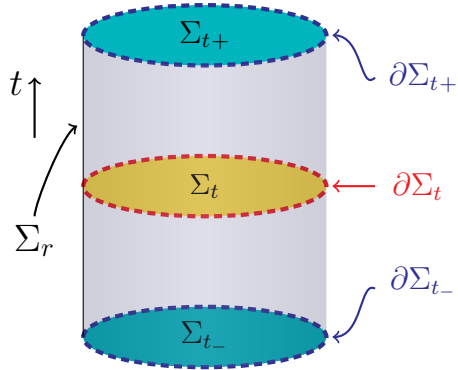	
	For studying conservation law in field theories it is natural to define charge $Q^{(t)}$ by integrating over a \emph{spacelike slice} ($t=const$ hypersurface $\Sigma_{t}$). However, one may note that the definition of charges over a spacelike hypersurface is not a necessity. Indeed, it is possible to define charge $Q^{(r)}$ by integrating over a \emph{timelike slice} ($r=const$ hypersurface $\Sigma_{r}$). From this perspective, temporal and spatial conservation are just a relation between charges over various time and spatial slices respectively (see figure \ref{fig:stcharges}).
    
As we mentioned several times, charges in gauge theories have a fascinating property: they are expressed in terms of a codimension-2 integral. In other words, in gauge theories, we can write the temporal (spatial) charge, $Q^{(t)}$ $(Q^{(r)})$, over the boundary of $\Sigma_t$ $(\Sigma_r)$ namely $\partial \Sigma_t$ $(\partial \Sigma_r)$.

This low-dimensional property of charges in gauge theories allows us to relate temporal and spatial conservation. Time evolution of $\partial\Sigma_{t}$ provides a natural timelike hypersurface $\Sigma_{r}=\mathbb{R} \times \partial \Sigma_{t} $ (see figure \ref{fig:stcharges}). Therefore, one may expect the temporal conservation of $Q^{(t)}$ and spatial conservation of $Q^{(r)}$ to be connected. In the interest of simplification, let us consider $\partial \Sigma_{t}$ as a two dimensional sphere $S^2$ (in contrast to the cylindrical boundary assumption in \ref{sec:Conservation-Memory}). Then its time evolution during $[t_{-},t_{+}]$ provides a timelike boundary $\Sigma_{r}=\mathbb{R} \times S^2$. As shown in figure \ref{fig:stcharges}, $\partial\Sigma_{r}=\partial\Sigma_{t_+}\cup \partial\Sigma_{t_-}$, obviously it implies $Q^{(r)}=Q^{(t)}\eval_{t_{+}}-Q^{(t)}\eval_{t_{-}}$. 
	
	By this inspiration, we have investigated Maxwell's theory in the presence of a timelike boundary and showed how the (generalized) Ampere-Maxwell {charge} appears as the surface charge associated with the standard $U(1)$ Maxwell gauge symmetry. 
	We also showed these surface charges obey a spatial conservation law. \footnote{ In this paper, we focus solely on the classical aspects of the subject. It is worth noting that extending this to quantum mechanics would require careful consideration of the positions of inserted operators when transitioning from a codimension-1 surface to a codimension-2 surface. We would like to express our gratitude to the anonymous referee for bringing this important point to our attention.}
	
    Furthermore, we defined the spatial memory field and its associated spatial memory effect as the same as the temporal one. In the context of Strominger's triangle proposal,  we showed how the generalized electric current (spatial charge) generates this memory field. Besides, we derived an explicit relation between spatial conservation and spatial memory field. In other words, 
	the spatial changes in the charges associated with residual gauge symmetry are encoded in the hard flux and the memory effect. More accurately, as indicated in the introduction,  here we do not have the notion of  IR physics
 similar to Strominger's IR triangle, but instead, we have sort of low momentum physics. 
	
	One may note that the conservation of $\Qla$ constrains electromagnetic fields generated by charges and current distributions. Constraints due to the conservation of $\Qlat$ over a radiating system have been partially studied in \cite{Seraj:2016jxi}. The notion of spatial conservation allows us to survey the implications of constraints even in the absence of radiation. In particular, it would be interesting to investigate the impact of \eqref{eq:DeltaIz} on currents and magnetic field in magneto-static. 
	
	In this paper, we only focus on the Maxwell theory in $1+3$ dimensions. But we expect these concepts to arise naturally in any theory with local symmetries. In this regard, studying non-Abelian gauge theories and diffeomorphism invariant theories of gravity will be interesting.
	\section*{Acknowledgement}
	We would like to thank A. Seraj, M.M. Sheikh-Jabbari, and  Y. Sobouti for helpful discussions and comments on the draft version.
	\appendix
\section{Symplectic Structure}\label{Symplectic Structure}
To construct the Poisson brackets, we follow the covariant phase space method \cite{Lee:1990nz}. In this regard, we start with the first variation of the Maxwell action \eqref{action} without the external electric source,
\begin{align}
	\delta S[A_\mu]=\int \dd^4x\qty(\partial_{\mu} F^{\mu \nu})\delta A_{\nu}+\int \dd^{4}x \;\partial_{\mu} \Theta^{\mu}\, ,
\end{align}
here the symplectic potential (boundary term) for the pure Maxwell theory is given by
 \begin{equation}
     \Theta^{\mu}=-F^{\mu\nu}\delta A_{\nu}\, ,
 \end{equation}
 where
 $\Theta^{i}=-E^{i}\;\delta A_{t}+\qty(\vb{B}\times\delta \vb{A})^{i}$ and
 $\Theta^{t}=-\vb{E}\cdot \delta \vb{A}$.
 We now define the symplectic two form on a $z=const$ surface as follows
 \begin{equation}
     \Omega=-\int_{\Sigma_z} \dd\Sigma_{\mu} \delta F^{\mu\nu} \wedge \delta A_{\nu}\, .
 \end{equation}
 Its explicit form in the cylindrical coordinate is given by
  \begin{equation}
     \Omega
     =-\int_{\Sigma_z} \dd t \dd{\sigma} \left(\delta F^{z t} \wedge \delta A_{t}+\delta F^{z s} \wedge \delta A_{s}+\delta F^{z \phi} \wedge \delta A_{\phi}\right)\, ,
 \end{equation}
where $\dd{\sigma}=s \dd s \dd\phi$. Let us work in the temporal gauge, $A_t=0$, and also we restrict ourselves to the static configurations of the solution space. In this case, we will have
 \begin{equation}
    \Omega=\int \dd t \;\omega\, , \hspace{1 cm} \omega:=-\int_{\Sigma_{zt}} \dd{\sigma} \left(\delta F^{z s} \wedge \delta A_{s}+\delta F^{z \phi} \wedge \delta A_{\phi}\right).
 \end{equation}
 In terms of the magnetic field, we find
 \begin{equation}
     \omega=-\int_{\Sigma_{zt}} s\dd s d\phi \left(\delta B_{\phi} \wedge \delta A_{s}-\delta B_{s} \wedge \delta A_{\phi}\right)\, .
 \end{equation}
 This result readily yields the following Poisson brackets
 \begin{equation}
    \begin{split}
        &\{A_s(s,\phi,z), B_{\phi}(s',\phi',z)\}= -\frac{1}{s}\delta (\phi-\phi')\delta (s-s')\, , \\
        &\{A_\phi(s,\phi,z), B_{s}(s',\phi',z)\}= \frac{1}{s}\delta (\phi-\phi')\delta (s-s')\, .
    \end{split}
 \end{equation}
 By using these canonical commutation relations, one can compute 
 \begin{equation}
  \pb{\Ilaz}{A_{s}(s,\phi,z)}=\partial_{s} \lambda(s,\phi,z)\, .
  \end{equation}
  This result simply shows that the charge $\Ilaz$ generates the gauge transformation on $A_{s}$.  The generalization of these computations to the dynamical cases is straightforward.
		\bibliographystyle{fullsort.bst}
		\bibliography{reference}
	\end{document}